# Predicting the connectivity of primate cortical networks from topological and spatial node properties

Luciano da F Costa[1], Marcus Kaiser[2,3] and Claus C Hilgetag*[4,5]

Address: [1]Instituto de Física de São Carlos, Universidade de São Paulo, Caixa Postal 369, São Carlos, SP, 13560-970, Brazil, [2]School of Computing Science, Newcastle University, Claremont Tower, Newcastle upon Tyne, NE1 7RU, UK, [3]Institute of Neuroscience, Henry Wellcome Building for Neuroecology, Newcastle University, Framlington Place, Newcastle upon Tyne, NE2 4HH, UK, [4]Jacobs University Bremen, School of Engineering and Science, Campus Ring 6, 28759 Bremen, Germany and [5]Boston University, Sargent College, Department of Health Sciences, 635 Commonwealth Ave, Boston, MA 02215, USA

Email: Luciano da F Costa - luciano@ifsc.usp.br; Marcus Kaiser - m.kaiser@newcastle.ac.uk; Claus C Hilgetag* - c.hilgetag@iu-bremen.de

* Corresponding author

## Abstract

**Background:** The organization of the connectivity between mammalian cortical areas has become a major subject of study, because of its important role in scaffolding the macroscopic aspects of animal behavior and intelligence. In this study we present a computational reconstruction approach to the problem of network organization, by considering the topological and spatial features of each area in the primate cerebral cortex as subsidy for the reconstruction of the global cortical network connectivity. Starting with all areas being disconnected, pairs of areas with similar sets of features are linked together, in an attempt to recover the original network structure.

**Results:** Inferring primate cortical connectivity from the properties of the nodes, remarkably good reconstructions of the global network organization could be obtained, with the topological features allowing slightly superior accuracy to the spatial ones. Analogous reconstruction attempts for the *C. elegans* neuronal network resulted in substantially poorer recovery, indicating that cortical area interconnections are relatively stronger related to the considered topological and spatial properties than neuronal projections in the nematode.

**Conclusion:** The close relationship between area-based features and global connectivity may hint on developmental rules and constraints for cortical networks. Particularly, differences between the predictions from topological and spatial properties, together with the poorer recovery resulting from spatial properties, indicate that the organization of cortical networks is not entirely determined by spatial constraints.

## Background

Scientific-technological advances over the last decades have produced ever-increasing experimental knowledge about brain organization and dynamics. In particular, modern anatomical techniques have provided extensive data on the interconnections of cerebral cortical areas in the brains of animals such as the cat or rat, or non-human primates such as the rhesus monkey. The intricate, non-random connectivity of cortical brain regions mediates the diverse and flexible sensory, cognitive and behavioral functions of the mammalian brain. However, the topological organization of these networks [1] as well as their spa-tial layout in the brain [2] are still incompletely understood. This is particularly apparent for the connectivity of the human cerebral cortex, which is largely unknown, due to experimental limitations [3].

A fundamental open problem in systems neuroscience is the relationship between specialized features of local nodes, such as areas of the cerebral cortex, and the global interaction and integration of these nodes in the neural networks. One aspect of this relationship concerns the question from which features of the local nodes structural connectivity between them might be predicted.

We address this question with the help of network analysis approaches [4]. Because cortical networks are typically complex, little insight can be obtained through their visualization alone. Therefore, useful objective and quantitative characterizations of complex networks ultimately rely on the estimation of a number of complementary measurements of their properties [5]. Network measurements typically provide information about specific topological or geographical features of the networks. For instance, the node degree provides a simple and valuable quantification of the intensity of connections between a specific node and the rest of the network. However, it says nothing about the origin or destinations of such connections. On the other hand, the clustering coefficient of a node provides an objective quantification of the degree in which the immediate neighbors of a node (nodes which can be reached directly without involving any intermediate nodes) are interconnected, but provides no information about the rest of the network. Because of the specificity and complementariness of typical network measurements, an essential question arises regarding what subsets of measurements are more complete, in the sense of allowing accurate, or at least reasonably approximate, reconstruction of the original network from its respective topological or geographical measurements. Remarkably, this question has been little explored in the complex networks literature (however, see [6] for an initial foray in this area).

It is important to note that the problem of network reconstruction from topological features is in a sense circular. Such features are derived from the complete connectivity of the network, so global connectivity may be inferred by taking itself into account. However, this is by no means a trivial task. For instance, guessing which nodes are specifically interconnected, based on measurements such as their degree or clustering coefficient, is almost invariably an impossible task. The exercise of trying to reconstruct the connections from a collection of topological measurements therefore provides an interesting new way to look at specific properties and structural organization of a complex network. For instance, in case the connectivity could be reasonably guessed from the node degree correlations, this would provide a key insight about its underlying organization.

We consider topological as well as spatial parameters, as biological networks, and brain networks in particular, are embedded in space. It is an interesting question to ask how the topological and spatial organization of these networks relates to each other. In particular, how do the topological and spatial features of individual nodes relate to the connectivity and layout of the whole network? Answers on these questions may inform current theories on the evolution and development of complex biological networks.

The Methods section of this article presents the adopted topological and spatial features and describes the reconstruction methodology based on similarity between sets of features. The analysis was applied to primate cortical brain connectivity (2,402 connections among 95 cortical areas of the Macaque monkey). In order to provide a comparative case, we also describe the application of the same methodology to *C. elegans* neuronal connectivity [Additional file 1].

## Results
### *Overview and community analysis*
Figure 1 presents a two-dimensional projection of the center of mass of the cortical areas together with their interconnections, obtained by principal component analysis [7]. This makes the three-dimensional organization of the cortical network accessible in two dimensions.

Figure 2 shows the frequency histograms of Euclidean distances between all pairs of nodes (a), number of existing edges with a given distance (b), and the ratio between the histograms in (b) and (a). Note that (a) represents the lengths of all potential links, while (b) shows the lengths of the actually existing connections between nodes. A series of interesting features can be inferred from these results. First, we see from (a) and (b) that the cortical network under analysis involves just a few pairs of edges which are close to one another (i.e. distances smaller than 10). This is a direct consequence of the fact that each cortical region has been represented in terms of its center of mass, therefore limiting the minimal distances between adjacent pairs. More interestingly, the ratios of existing edges per possible pairs in histogram (b) clearly indicate (by the decaying profile of this histogram as the distance increases) that the further away a pair of regions is, the less likely their interconnection.

Given a network, it is often the case that a subset of its nodes is more interconnected with one another than with the remainder of the network. Such a subset of nodes, together with the respective interconnections, is called a *community* inside the original network. The intensity of the separation between the community and the rest of the network can be quantified in terms of its *modularity index*, which varies between 0 and 1 [8]. In order to determine the main communities in the cortical system, we applied Newman's spectral method [8] and obtained the two regions identified in Figure 3. This approach to community detection is based on rewriting the modularity function of the network in terms of matrices, so that the best partition in two communities can be obtained in terms of spectral analysis of those matrices. Further subdivision of such regions was unjustified because of the low modularity values obtained for such subdivisions. Our approach helped to ensure that the subsequent analysis was not biased by gaps between different datasets describing the cortical network (cf. Methods, section 'Neural network data').

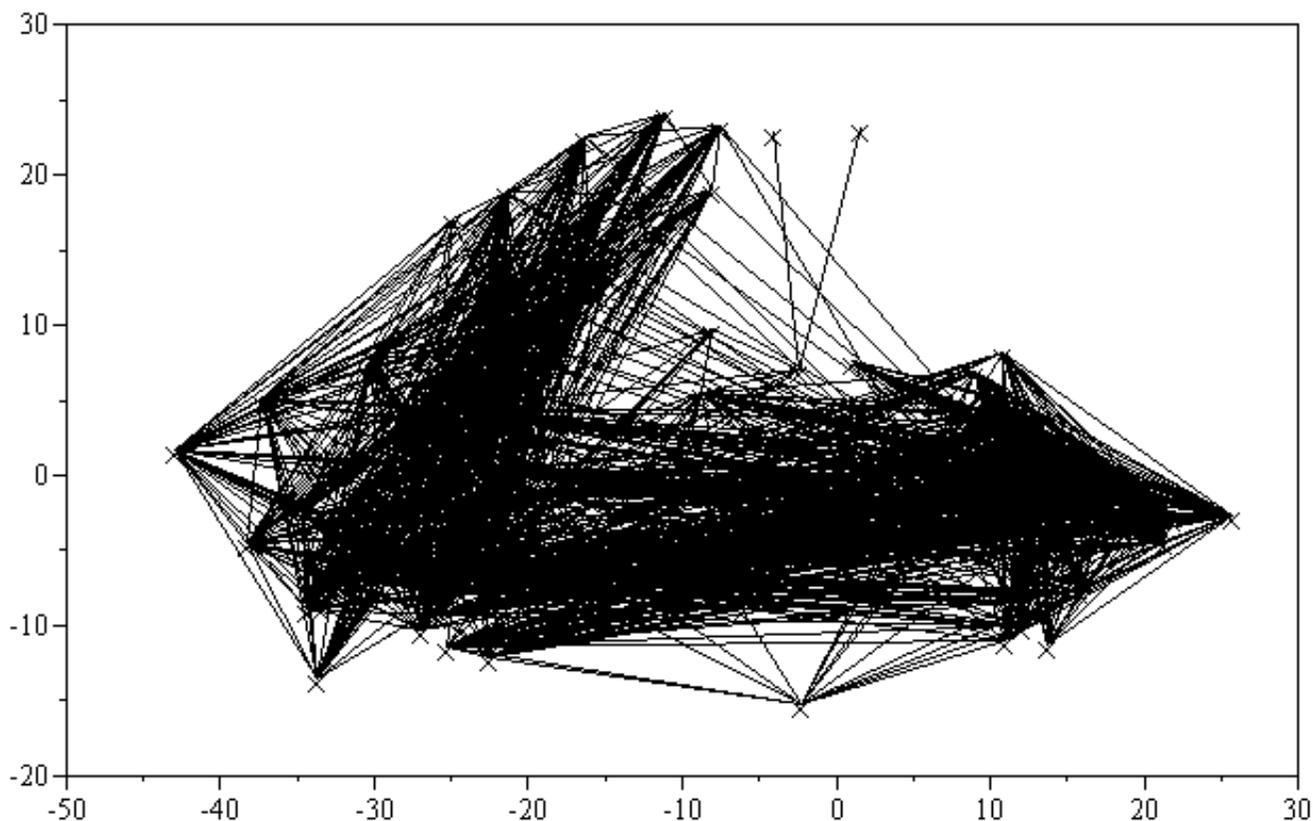

**Figure 1**
Two-dimensional projection of the network of 95 cortical area-nodes at the centers of mass of the cortical regions, obtained through principal component analysis and for the complete set of their existing interconnections.

Communities 1 and 2 were of comparable size and included $N_1 = 44$ and $N_2 = 51$ nodes, respectively, and $E_1 = 1326$ and $E_2 = 1280$ directed edges. The clustering coefficients obtained for the two identified communities were found to be equal to 0.52 and 0.68, respectively. This might be explained by the higher number of connections within communities. Whereas the global edge density of the network was 0.17, the densities within the communities were 0.50 and 0.66.

*Topological characterization*
Now we focus our attention on the analysis of the local node properties and connectivity of these two communities. Figure 4 presents the histograms for node degree, clustering coefficient, and matching index with respect to the two identified communities. Similar histograms were obtained for most measurements, except the node degree, which resulted markedly different in each community, being more evenly distributed in the case of community 2. Clustering coefficients of individual nodes in both communities were above 0.5. In addition, the average clustering coefficient was both above the global density as well as above the edge density within the respective community. The probability of average shortest path distances appears to decay with the distance. The matching index within the communities was between 0.5 and 0.6 as would be expected from nodes within the same community. Note, however, that some nodes had a matching index below 0.3 indicating outlier nodes.

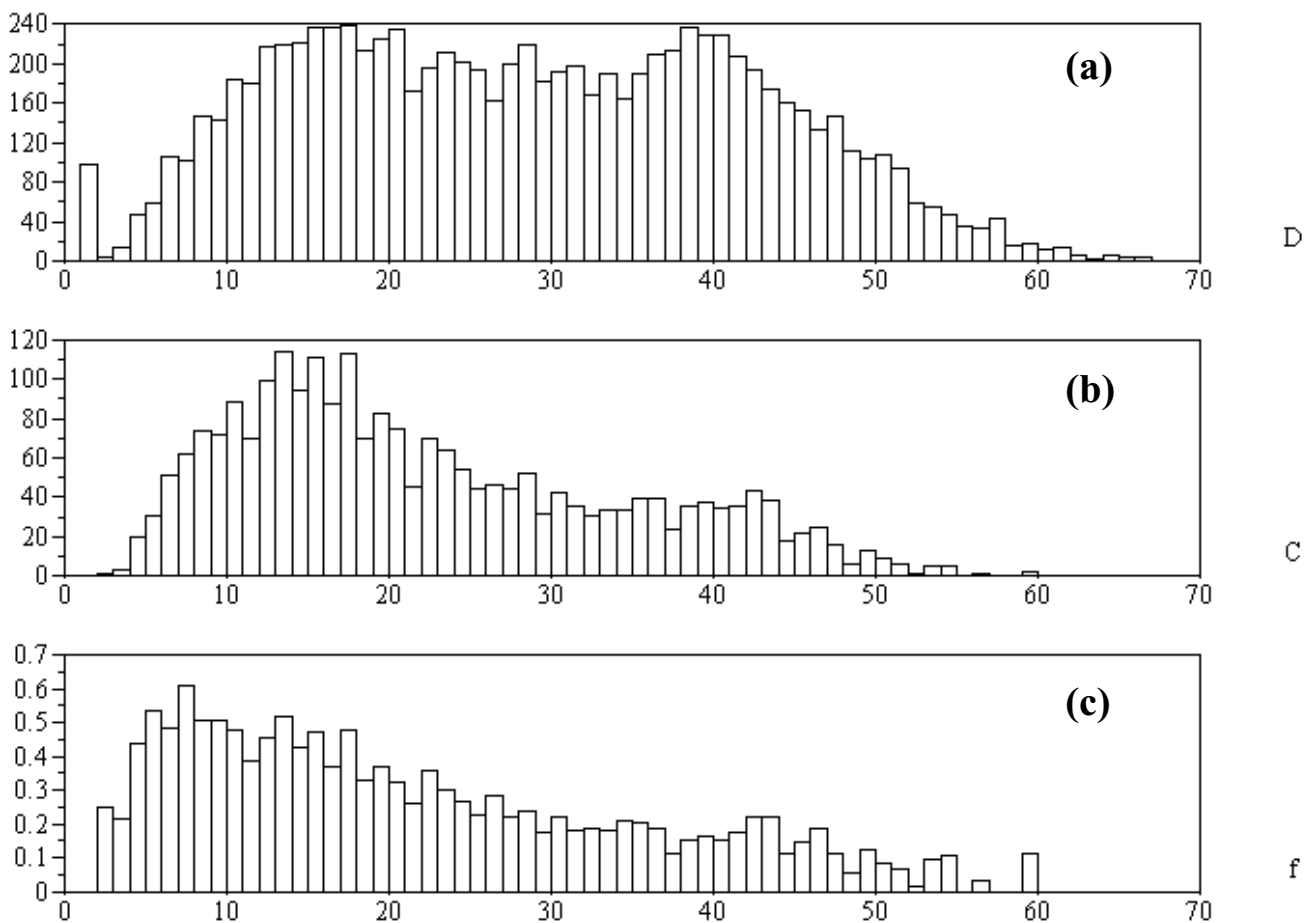

**Figure 2**
Spatial distribution of network connections. The histograms of Euclidean distances between all pairs of edges (a), number of existing edges with respective distance (b) and the ratio between (b) and (a).

*Spatial characterization*
Figure 5 shows the histograms for average topological distance, and average effective distance, and area obtained for the two communities. It is clear from these results that similar averages characterize the measurements in each community, while the respective distribution varies markedly.

An important issue to be considered while adopting several measurements is the quantification of possible relationships between them, which can be indicated by the Pearson correlation coefficients for all pairwise combinations of measures. The Pearson coefficients calculated independently for the topological and spatial measurements are given in Tables 1 and 2. It follows from the results in Table 1 that the node degree is strongly correlated with the matching index, while exhibiting moderate anticorrelation with the average shortest path distance. As could be expected, the local density was found to be weakly anticorrelated with the area size (Table 2).

Except for the strong correlation between the node degree and matching index, all other pairs of measurements were unremarkable, supporting the complementariness of the adopted sets of features.

*Comparison between original and reconstructed networks*
Table 3 gives the expected average ratios of correct ones and zeroes, as well as their respective geometrical averages, for the two principal communities in the cortical networks.

We performed an exhaustive search while taking into account all 1-by-1, 2-by-2 and 3-by-3 combinations of each of the two types of considered measurements for a whole sequence of threshold values *T* (ranging from 0.1 to 7 in steps of 0.1) in order to identify those combinations producing reconstructed networks which are most similar to the original network *G*. Table 4 presents the combinations of measurements and respective geometrical average of $R_1$ and $R_2$ with respect to the two main cortical communities considered in this work. It is clear from this table that the best synthesized networks were obtained by the matching index for the first community and the combination of (clustering coefficient, matching index) for the second community.

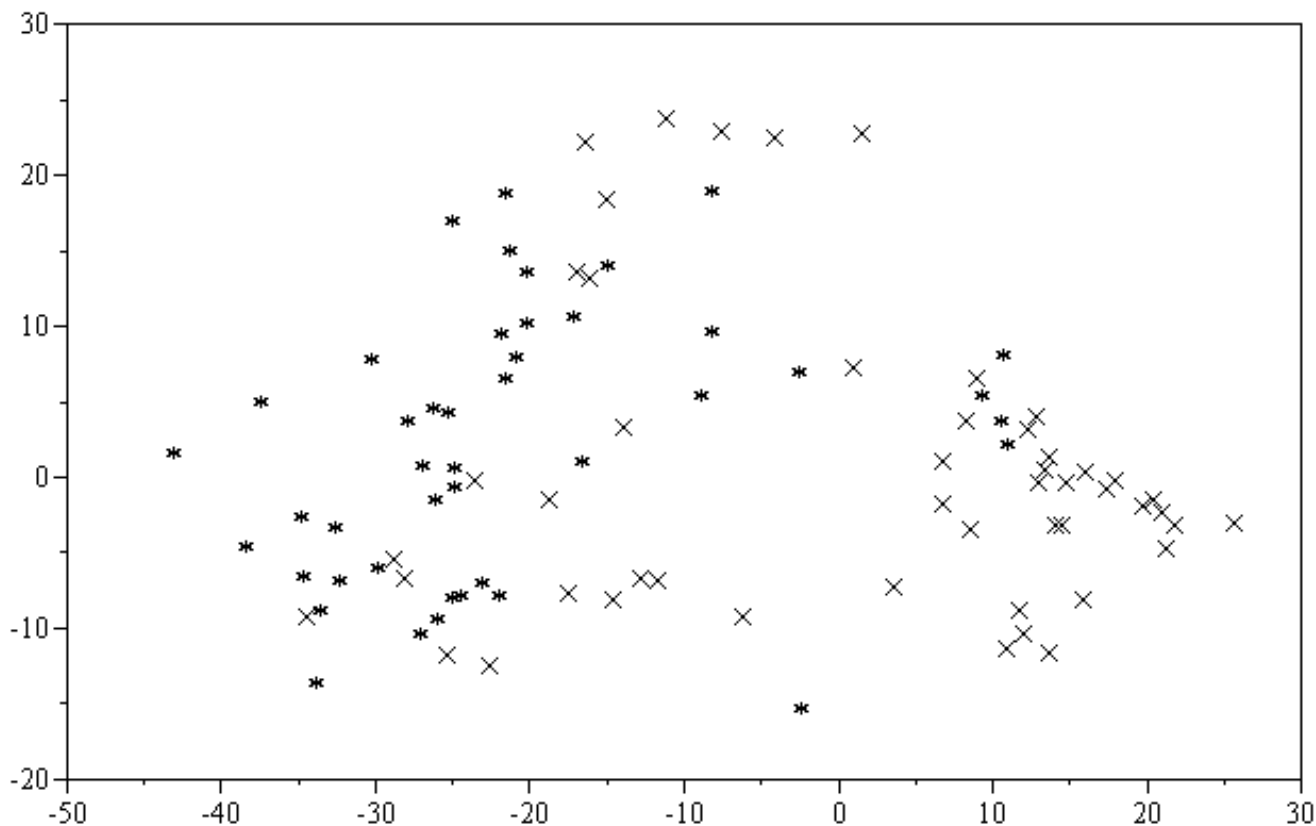

**Figure 3**
The two identified communities represented in terms of the center of mass of each cortical region. Communities 1 and 2 are identified respectively by '×' and '*'.

Figure 6 shows the adjacency matrices of the original communities (a,b) and those of the respective most similar networks (c,d) obtained by considering the topological properties at each node. Remarkably, the networks constructed on the basis of the combinations of measurements appeared reasonably similar to the respective original networks.

The qualities of the reconstructions obtained by considering the 4 spatial features are given in Table 5. The best reconstruction of communities 1 and 2 were obtained for the local density and local density/area size, respectively.

In order to investigate how the combinations of topological and spatial features perform with respect to the network reconstruction, we also considered hybrid combinations between the two topological (i.e. clustering coefficient and matching index) and the two spatial (i.e. local density and area size) features which were found to produce the best results in Table 4 and 5, respectively. The results are given in Table 6. The best reconstruction of community 1 was obtained as before by considering only the matching index. However, a small improvement was observed for community 2 as allowed by the combination between the two topological features plus the area size. The respective network reconstruction is not shown as it is very close to that obtained for the two topological features

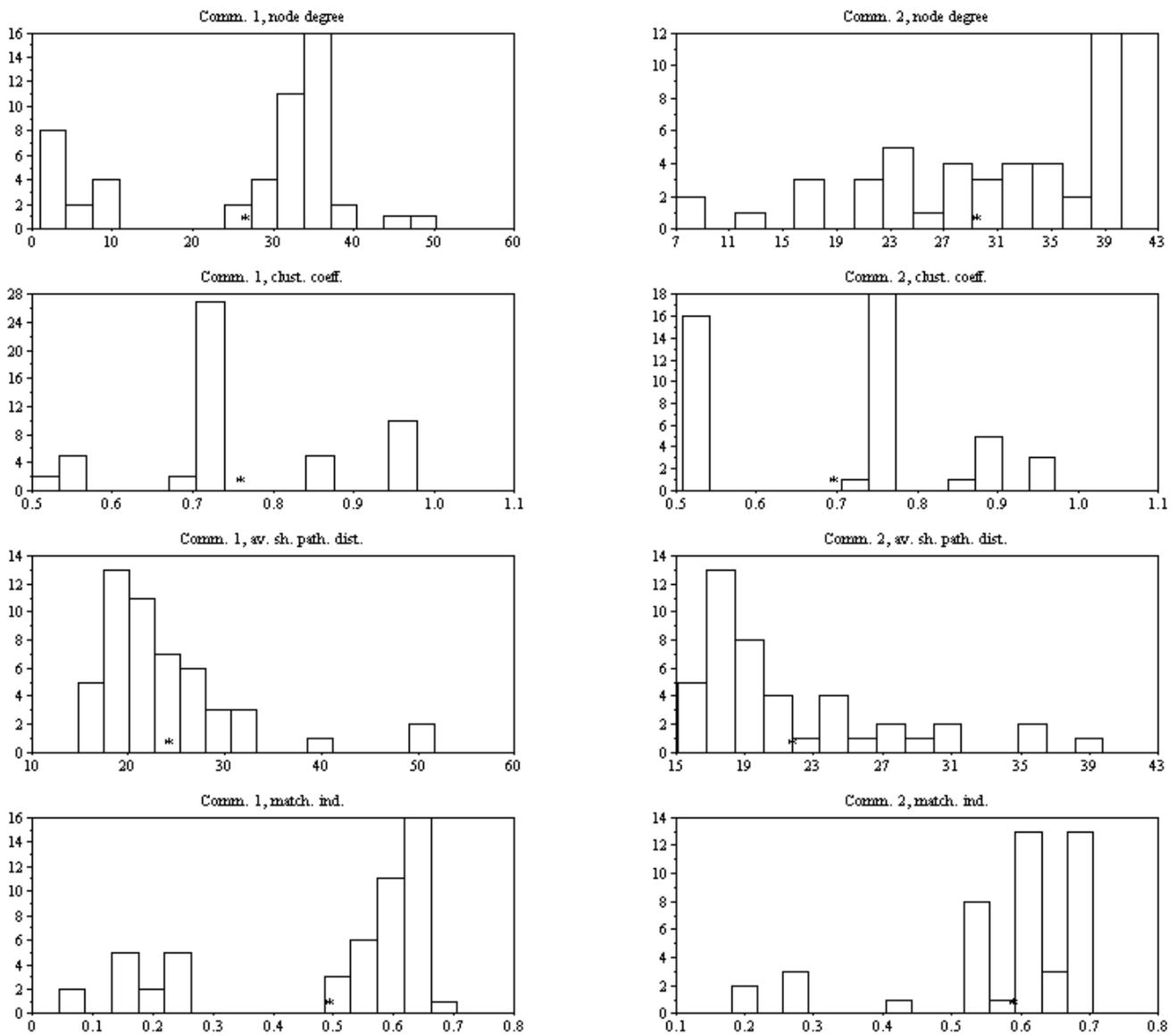

#### Figure 4
Histograms of node degree, clustering coefficient, average distance, average topological distance, and matching index for community 1 (left-hand side column) and community 2 (right-hand side column). The asterisks identify the average of each measure. The histogram distributions of the two communities differed markedly for the node degree, however, the average degrees of the communities were very similar for all measures.

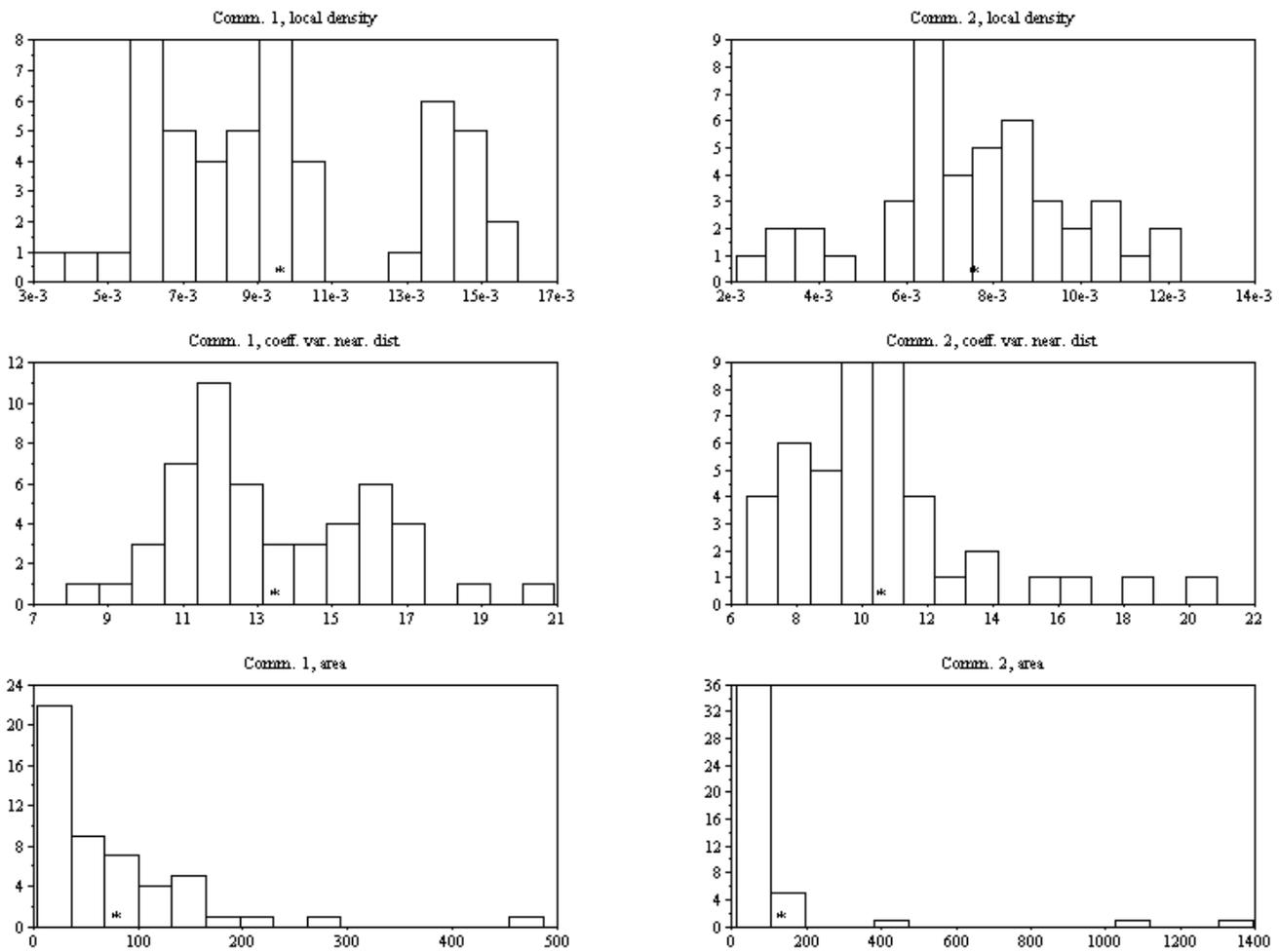

#### Figure 5
Histograms of local density, coefficient of variation, and area obtained for the two communities. The three Cartesian coordinates are not shown because they are not invariant with position and rotation and can be readily inferred from the projection in Figure 3. The asterisks show the average of each measure. Although the average of the measurements for each respective community tends to be similar, the distributions in all histograms are markedly different.

**Table 1: Correlations of topological node-based measures.**

|  | Node degree | Clustering coefficient | Avg. shortest path distance | Matching Index |
| --- | --- | --- | --- | --- |
| Node degree | 1.00 | --- | --- | --- |
| Clustering coefficient | 0.36** | 1.00 | --- | --- |
| Avg. shortest path distance | -0.46** | -0.28** | 1.00 | --- |
| Matching index | 0.94** | 0.34** | 0.46** | 1.00 |

Pearson correlation coefficients were obtained for all pairs of considered topological measurements (** indicates that correlation is significant at the 0.01 level, two-tailed).

**Table 2: Correlations of spatial node-based measures.**

|  | Local density | Coeff. variation nearest distances | Area |
|---|---|---|---|
| Local density | 1.00 | --- | --- |
| Coeff. Variation nearest distances | 0.22* | 1.00 | --- |
| Area | -0.10 | 0.11 | 1.00 |

Pearson correlation coefficients were obtained for all pairs of considered spatial measurements (* indicates that correlation is significant at the 0.05 level, two-tailed).

(i.e. clustering coefficient and matching index, see Figure 5b).

Figure 7 shows the original and reconstructed matrices considering the 4 spatial features.

Interestingly, a comparison between the adjacency matrices in Figure 6 and 7 immediately shows that the networks inferred on the basis of the measurements of topological properties at each node reproduced the original connectivity better than networks constructed by the consideration of the spatial properties.

It is quite surprising that such good reconstructions of the original matrices could be obtained by considering relatively simple topological and spatial features. Table 7 summarizes the comparison between the original and reconstructed communities while considering topology and geometry. It is clear from this table that reasonable reconstruction can be obtained for the global organization of the cortical communities based on their local node properties. The geometrical averages also indicate that the two communities are possibly organized according to different topological and spatial influences, with community 1 being more strongly constrained by the adopted measurements.

*Predicting unknown connections*

Connections which have not yet been tested in tract tracing studies were so far treated as absent in this study. This is due to the fact that only one of the three compilations contributing to the present dataset distinguished between absent and unknown connections. For this compilation [9], which forms a 32 × 32 area subgraph, we reviewed reconstructed networks in the light of whether they were able to predict previously unknown connections. In this analysis, one area, VP, had to be excluded from the original matrix due to its unknown spatial position. For the combination of the best two topologic and two spatial measures, 111 currently unknown projections were predicted to exist, and 174 connections were predicted to be absent, yielding a realistic ratio for predicted existing connections of 39%, out of all unknown connections. The predicted projections are shown as yellow fields in the reconstructed subgraph matrix in Fig. 8. The figure also indicates mismatches (red fields) between the original and reconstructed matrices, either existing connections that were left out of the reconstructed matrix (90 cases) or absent connections filled in the reconstructed matrix (106 cases). Most entries (in green fields), however, were confirmed to exist (207 cases) or to be absent (212 cases).

We also explored the impact of the potential existence of the currently unknown connections, by creating two additional simulated versions of the 31 × 31 area subgraph matrix, in which (a) all unknown connections were assumed to exist ('full' version), (b) 31% of the unknown connections were assumed to exist (this reflects the average edge density in cortical networks, 'relative' version). Reasonable reconstructions were obtained in all these three cases, as demonstrated by the respective Hamming distances and geometrical average errors (Table 8).

## Discussion

We have explored the role of local topological and spatial features in determining cortical connectivity. Topological features had been analyzed before [6] with a measure similar to the matching index used here as a predictor of primate visual cortex connectivity. Previous studies were also applying the notion of neighborhood as a predictor of

**Table 3: Expected ratios of correct ones and zeroes, and respective geometrical average.**

| **Network** | $R_1 = r_1$ | $R_0 = r_0$ | $\sqrt{R_1 R_0}$ |
|---|---|---|---|
| **Cortical community 1** | 0.68 | 0.32 | 0.46 |
| **Cortical community 2** | 0.49 | 0.51 | 0.50 |

Theoretically expected average ratios of correct ones and zeroes, as well as their respective geometrical averages, obtained for the two cortical communities.

**Table 4: Network reconstruction from individual and combined topological node measures.**

|   | Community1 | Community2 |
|---|---|---|
| **Measurements** | $\sqrt{R_1 R_0}$ | $\sqrt{R_1 R_0}$ |
| 1 | 0.797 | 0.647 |
| 2 | 0.563 | 0.577 |
| 3 | 0.539 | 0.611 |
| 4 | **_0.810_** | 0.664 |
| 1, 2 | 0.637 | 0.706 |
| 1, 3 | 0.770 | 0.670 |
| 1, 4 | 0.802 | 0.663 |
| 2, 3 | 0.532 | 0.638 |
| 2, 4 | 0.704 | **_0.750_** |
| 3, 4 | 0.782 | 0.678 |
| 1, 2, 3 | 0.688 | 0.694 |
| 1, 2, 4 | 0.782 | 0.720 |
| 1, 3, 4 | 0.800 | 0.678 |
| 2, 3, 4 | 0.689 | 0.703 |

Geometrical averages of the connectivity estimation obtained for the two communities while considering the 14 combinations of measurements listed in the first column. The best combinations for communities 1 and 2 were respectively the matching index (4) and the pair of measurements involving the clustering coefficient (2) and matching index (4). The configurations leading to the best matches have been emphasized.

connectivity which suggests that spatially close regions tend be connected by fiber tracts [10]. In this article, we have expanded such notions by testing the relative impact of several topological and spatial constraints on neural network organization.

In general, a small number of local features is sufficient for predicting connections between regions. In the case of the topological features, the matching index represented the most effective individual feature for reconstruction of both communities, while the best selection for community 2 also required the clustering coefficient. This result substantiates the particular role of this feature for cortical organization [11] and means that cortical areas which have similar inputs and outputs also tend to be connected with each other. The best reconstructions obtained from spatial features were obtained by considering the local density for both communities. The area size was also required for the best reconstruction of community 2. These results suggest that regions with similar local densities tend to connect to one another. In the case of community 2, region interconnections also appear to favor similar area size.

Concerning single features for the prediction of connections, topological features led to a better estimation than spatial features. This may be partly explained by the fact that topological node features by their definition are indirectly linked to global network organization, as mentioned previously. It is, however, surprising that the 'purest' spatial parameter (parameter 8: area coordinates, which expresses the proximity between areas) did not result in a strong prediction for connectivity, as spatial distance has been previously put forward as an important factor in primate cortical connectivity [10]. This can be explained by the existence of a significant number of long-range connections in cortical networks, resulting from the fact that some regions are part of a network cluster but nonetheless spatially distant. In these cases, such as the frontal eye field being spatially distant from the rest of the visual cortex, spatial proximity would not predict a connection. Indeed, there exists a significant proportion of long-distance connections in biological neural networks [12] which ensures a low number of processing steps across these systems [2].

Since previous tract tracing studies have focused on the visual cortex, there might exist additional connections mainly within and between motor, auditory, and somatosensory cortices. As demonstrated for a smaller subgraph of the primate cortical network, our reconstruction approach could be used to guide future experimental studies, by deriving hypotheses about currently unknown projections which would be expected to exist or be absent. The analysis of different versions of this subgraph, with varying proportions of unknown connections assumed to exist, also demonstrated that the principal conclusions of this study do not depend on the number of currently unknown connections which may be discovered in the future.

An earlier analysis of the relationship between the surface size of cortical areas and the number of projections they send or receive found no significant correlation between these parameters [13]. The present analysis suggests that area size may be a factor contributing to the prediction of connections, after all (Results, section 'Comparison between original and reconstructed networks'). Thus, perhaps what matters is not the absolute area size, but the *matched* size of the connected regions.

For the feature analysis we transformed unidirectional projections into bidirectional connections. This resulted in 3,044 directed edges compared to the original 2,402 directed edges. This step was necessary as the reconstruction based on spatial distance depends on the Euclidean distance which is symmetric in both directions. It may be an interesting task for the future to repeat the topological analyses based on unidirectional measures.

The observed relationships between local node properties and global connectivity may hint on developmental rules. As the reconstruction approach worked well for the primate network, but not for the neuronal connections in *C.*

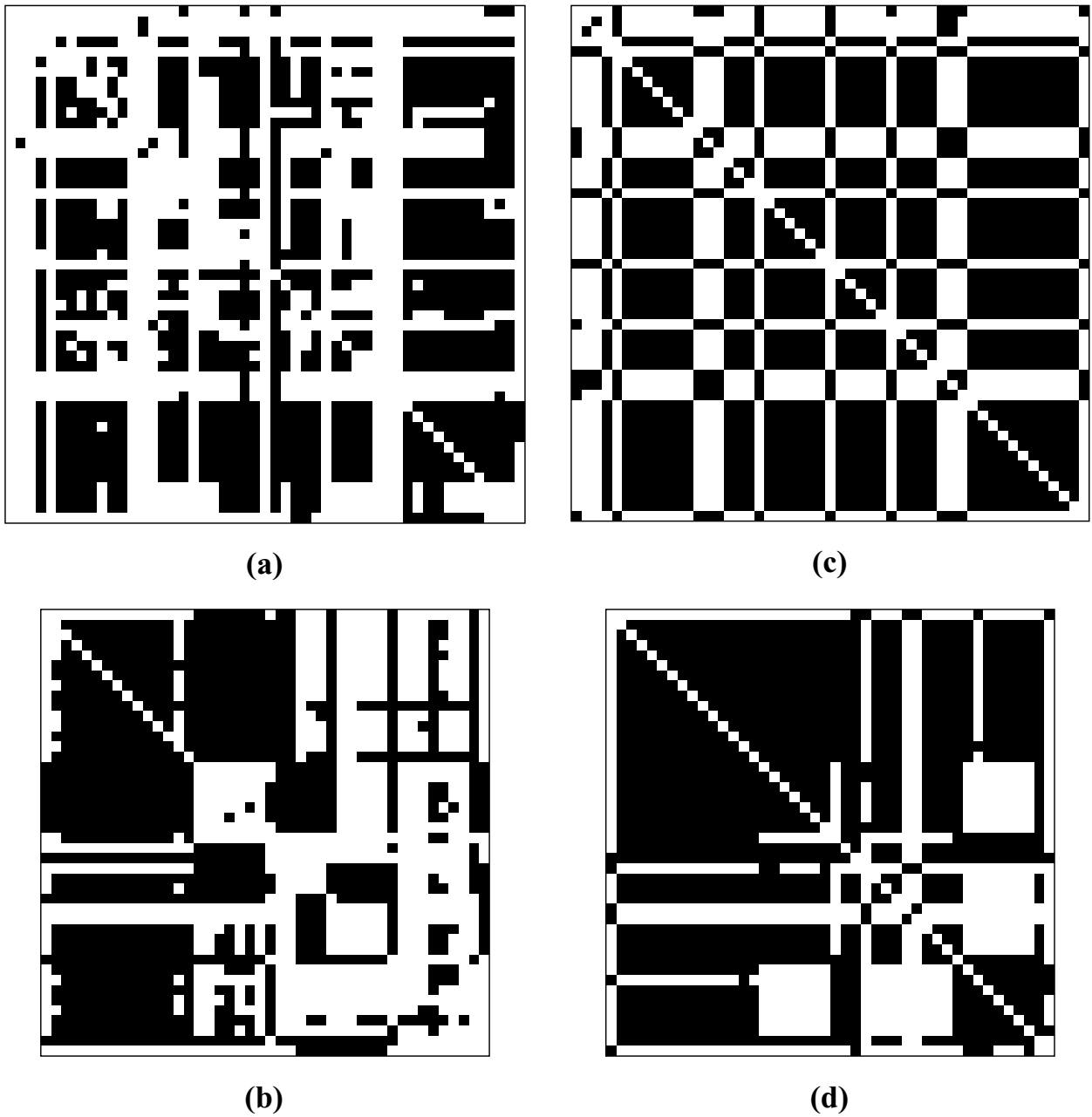

**Figure 6**
Adjacency matrices of the original communities (a,b) and those of the respectively most similar reconstructed networks (c,d) considering topological features. The reconstructions in this figure correspond to the highlighted configurations in Table 4. Black points indicate presence of connections.

**Table 5: Network reconstruction from individual and combined spatial node measures.**

| Measurements | Community1 $\sqrt{R_1 R_0}$ | Community2 $\sqrt{R_1 R_0}$ |
| --- | --- | --- |
| 5 | **_0.725_** | 0.612 |
| 6 | 0.471 | 0.587 |
| 7 | 0.498 | 0.554 |
| 8 | 0.503 | 0.552 |
| 5, 6 | 0.602 | 0.635 |
| 5, 7 | 0.654 | **_0.643_** |
| 5, 8 | 0.503 | 0.554 |
| 6, 7 | 0.451 | 0.621 |
| 6, 8 | 0.503 | 0.554 |
| 7, 8 | 0.504 | 0.555 |
| 5, 6, 7 | 0.574 | 0.638 |
| 5, 6, 8 | 0.503 | 0.554 |
| 5, 7, 8 | 0.504 | 0.556 |
| 6, 7, 8 | 0.504 | 0.554 |

Geometrical averages of connectivity estimation obtained for the two cortical communities while considering the 14 combinations of measurements listed in the first column. The best combinations for communities 1 and 2 were respectively the local density (5) and the pair of measurements including the local density (5) and cortical area (7). The configurations leading to the best matches have been emphasized.

**Table 6: Network reconstruction from combinations of topological and spatial node measures.**

| Measurements | Community1 $\sqrt{R_1 R_0}$ | Community2 $\sqrt{R_1 R_0}$ |
| --- | --- | --- |
| 2 | 0.563 | 0.577 |
| 4 | **_0.810_** | 0.664 |
| 5 | 0.727 | 0.612 |
| 7 | 0.498 | 0.584 |
| 2, 4 | 0.710 | 0.750 |
| 2, 5 | 0.590 | 0.650 |
| 2, 7 | 0.495 | 0.634 |
| 4, 5 | 0.790 | 0.685 |
| 4, 7 | 0.753 | 0.727 |
| 5, 7 | 0.654 | 0.643 |
| 2, 4, 5 | 0.753 | 0.721 |
| 2, 4, 7 | 0.688 | **_0.775_** |
| 2, 5, 7 | 0.595 | 0.673 |
| 4, 5, 7 | 0.753 | 0.708 |

Geometrical averages of connectivity estimation obtained for the two cortical communities while considering the 14 combinations of measurements listed in the first column. The best combinations for communities 1 and 2 were respectively the local density (5) and the pair of measurements including the local density (5) and cortical area (7). The configurations leading to the best matches have been emphasized.

*elegans* [see Additional file 1], it appears that the organization of neural networks is subject to different constraints in these two systems. One possibility is that the neuronal network of *C. elegans*, which is identical in each organism, could be largely determined by genetic factors [14] which may prescribe a specific connectivity independent of simple topologic or spatial rules. For larger neural systems, however, it may be impossible to encode the entire connectivity between cortical regions within the genome, resulting in a larger contribution of spatial and topologic constraints in the self-organization of systems connectivity.

Although the connectivity in non-human primates such as the macaque monkey is relatively well known, there is still only little information available about human connectivity. New methods such as diffusion tensor imaging [15] or post-mortem tract tracing [16] are applied to human brains but are still hindered by severe experimental limitations. It is our hope that the topological and spatial features reported in this study may complement and steer the current experimental approaches. These features could provide a basis for assessing the reliability of fiber tract predictions that are based on non-invasive methods.

## Conclusion
The reconstruction of neural connectivity from local node properties offers insights into constraints of network organization. In particular, it suggests that neuronal networks in *C. elegans* and neural networks in the primate cerebral cortex developed under different constraints, and that the layout of primate cortical brain networks is not entirely determined by spatial properties.

## Methods
### Neural network data
We analyzed the organization of 2,402 projections among 95 cortical areas and sub-areas of the primate (Macaque monkey) brain. The connectivity data were retrieved from CoCoMac ([17,18]) and are based on three extensive neuroanatomical compilations [9,19,20] that collectively cover large parts of the cerebral cortex. In the database, reported projections between cortical areas are based on anatomical tract tracing studies where dyes were injected into one cortical area, and anterograde or retrograde transport of the dye indicated target or source areas for projection fibers. Spatial positions of cortical areas were estimated from surface parceling using the CARET software http://brainmap.wustl.edu/caret. The spatial positions of areas were calculated as the average surface coordinate (or center of mass) of the three-dimensional extension of an area (cf. [2]). While this cortical data set is more extensive than those used in previous studies, it may still be partially incomplete, particularly for connections of motor, auditory and somatosensory areas. The restric-

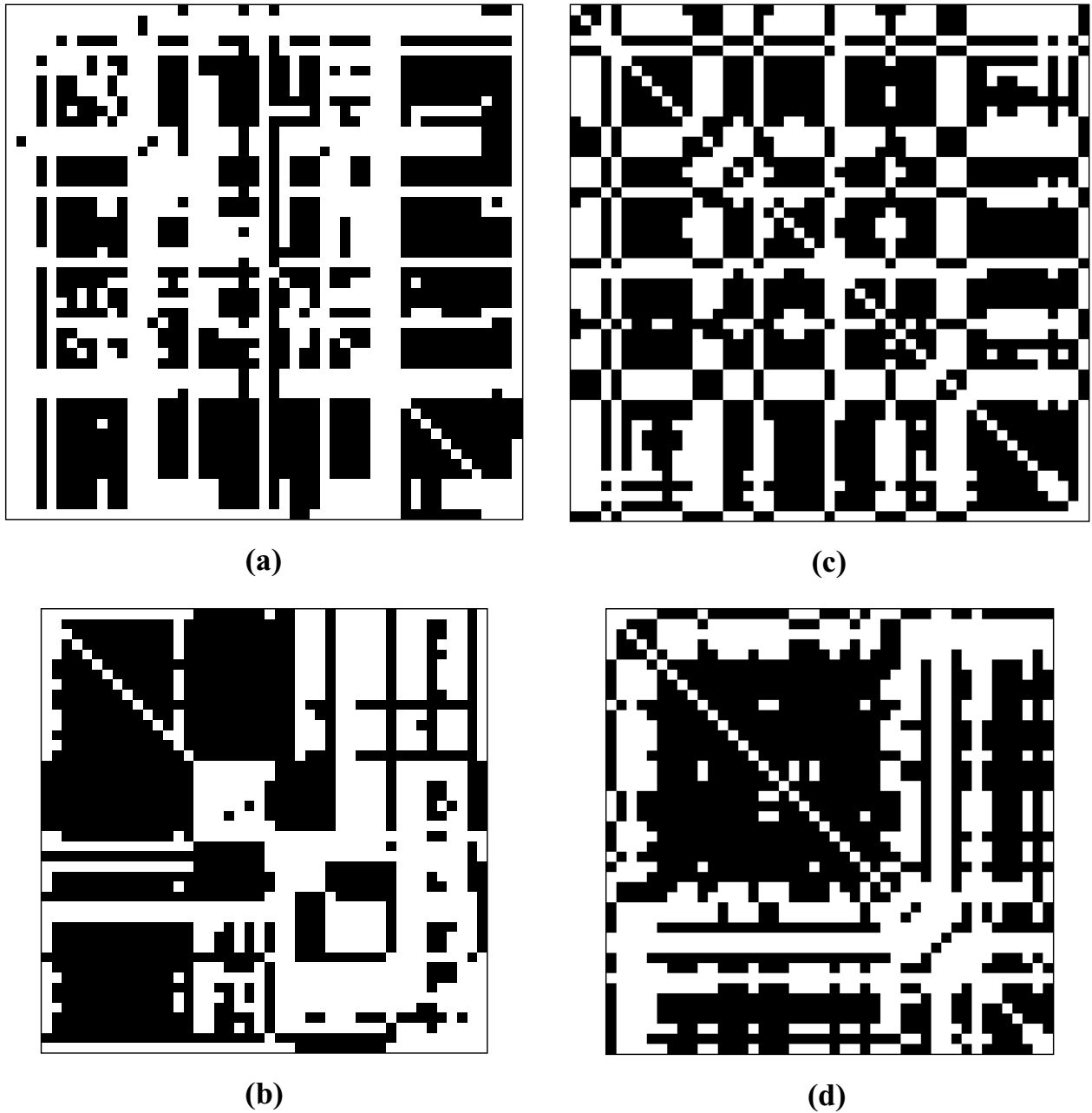

**Figure 7**
Adjacency matrices obtained for the two networks constructed based on the 4 spatial features. Original communities (a,b) and the respectively most similar reconstructed networks (c,d). Black points indicate presence of connections.

**Table 7: Comparison of network reconstructions with random benchmarks.**

| Network | Hamm. Dist. | $R_1$ | $R_0$ | $\sqrt{R_1 R_0}$ Random | $\sqrt{R_1 R_0}$ |
|---|---|---|---|---|---|
| Cortex/Comm. 1 Features reconstr. | 1486 | 0.793 | 0.831 | 0.46 | **0.810** |
| Cortex/Comm. 1 Distance reconstr. | 711 | 0.664 | 0.792 | 0.46 | **0.725** |
| Cortex/Comm. 1 Mixed feats. reconstr. | 1486 | 0.793 | 0.831 | 0.46 | **0.810** |
| Cortex/Comm. 2 Features reconstr. | 898 | 0.641 | 0.921 | 0.50 | **0.750** |
| Cortex/Comm. 2 Distance reconstr. | 764 | 0.513 | 0.807 | 0.50 | **0.643** |
| Cortex/Comm. 2 Mixed feats. reconstr. | 520 | 0.954 | 0.628 | 0.50 | **0.775** |

Overview of measurements comparing the original and reconstructed networks: ratio of overall matches, percentage of correct zeros, percentage of correct ones, and geometrical averages between the two latter percentages expected for random comparisons and obtained from the considered experiments. It is apparent that the topological and spatial reconstructions of the two cortical communities have quality substantially superior to the random reference.

tion arose from the fact that only studies could be used for which a parcellation scheme with spatial coordinates existed in CARET. In order to avoid potential artifacts associated with the segregation of the data in available reports, we first performed a community analysis of the cortical network and then analyzed the identified two communities separately (Results, section 'Overview and community analysis').

For comparison, we also analyzed two-dimensional spatial representations of the rostral neuronal network (131 neurons, 764 connections) of the nematode *C. elegans* [see Additional file 1]. Spatial two-dimensional positions (in the lateral plane), representing the position of the soma of individual neurons in *C. elegans*, were provided by Y. Choe [21]. Neuronal connectivity was obtained from [22]. The dataset was slightly modified as described in detail in [2].

The cortical as well as the *C. elegans* datasets are available at [23].

*Graph-theoretical representation*

The connections between cortical regions can be represented and understood as a graph, eg, [10,24], or complex network, eg, [1]. More specifically, the $N = 95$ cortical areas considered in this work are represented as nodes while the existing connections between such nodes are expressed in terms of edges. More formally, the cortical network is represented in terms of its *adjacency matrix* $K$, with dimension $N \times N$, with the presence of a connection extending from node $j$ to node $i$ being indicated as $K(i, j) = 1$. The adjacency matrix only gives information about whether a connection between two nodes exists in the network; in particular, it contains only topological information about the network and is not related to the colloquial meaning of adjacent as spatially nearby. A non-directed version $K_{non}$ of the adjacency matrix $K$ can be obtained as

$$\begin{cases} K_{non}(i,j) = 1 & \text{if } (K(i,j) + K(j,i)) > 0 \\ K_{non}(i,j) = 0 & \text{otherwise} \end{cases} \quad (1)$$

Because we also have information about the spatial position of each cortical region, it is possible to construct a *distance matrix* $D$ such that $D(i, j)$ represents the Euclidean distance between nodes $i$ and $j$. Note that both matrices $K_{non}$ and $D$ are symmetric by construction, i.e.: $K_{non}(i,j) = K_{non}(j,i)$ and $D(i,j) = D(j,i)$ for any $i$ and $j$. It is possible to calculate a series of measurements from matrices $K$, $K_{non}$ and $D$ in order to characterize the topological and spatial properties of the original network. For these measurements, we used the symmetric topological adjacency matrix $K_{non}$ to be comparable with the symmetric spatial distance matrix $D$. While such measurements are often performed for the network as a whole, here we focus on local measurements obtained for each network node.

*Network characterization indices*

The following 8 node-based measurements (4 topological and 4 spatial) were considered in the analysis.

*Feature 1 (Topological) – Node degree*

This simple but informative measurement quantifies the number of edges attached to a node. In the case of non-directed networks, the node degree of node $i$ can be calculated from the respective adjacency matrix as:

|     | V1 | V2 | V3 | PIP | V3A | V4 | PO | MT | V4t | DP | VOT | LIP | VIP | MIP | MDP | MSTd | MSTl | PITd | PITv | 7a | STPp | CITd | CITv | STPa | AITv | FEF | TF | 46 | FST | TH | AITd |
|-----|----|----|----|-----|-----|----|----|----|-----|----|-----|-----|-----|-----|-----|------|------|------|------|----|------|------|------|------|------|-----|----|----|-----|----|------|
| V1  |    | 0  | 0  | 0   | 0   | 1  | 0  | 0  | 0   | 0  | 0   | 0   | 0   | 0   | 0   | 0    | 0    | 1    | 1    | 1  | 1    | 1    | 1    | 1    | 0    | 1   | 1  | 0  | 1   | 1  | 1    |
| V2  | 0  |    | 1  | 0   | 1   | 1  | 1  | 1  | 1   | 1  | 0   | 1   | 1   | 0   | 0   | 1    | 1    | 0    | 1    | 1  | 0    | 1    | 0    | 0    | 1    | 1   | 1  | 1  | 1   | 0  | 0    |
| V3  | 0  | 1  |    | 0   | 1   | 1  | 1  | 1  | 1   | 1  | 0   | 1   | 1   | 0   | 0   | 1    | 1    | 0    | 1    | 1  | 0    | 0    | 0    | 0    | 0    | 1   | 1  | 1  | 1   | 0  | 0    |
| PIP | 0  | 0  | 0  |     | 0   | 0  | 0  | 0  | 0   | 0  | 0   | 0   | 0   | 0   | 0   | 0    | 0    | 0    | 0    | 0  | 0    | 0    | 0    | 0    | 0    | 0   | 0  | 0  | 0   | 0  | 0    |
| V3A | 0  | 1  | 1  | 0   |     | 1  | 0  | 1  | 1   | 1  | 0   | 1   | 1   | 0   | 0   | 1    | 1    | 0    | 1    | 1  | 0    | 0    | 0    | 0    | 0    | 1   | 1  | 1  | 1   | 0  | 0    |
| V4  | 1  | 1  | 1  | 0   | 1   |    | 0  | 1  | 1   | 1  | 0   | 1   | 1   | 0   | 0   | 1    | 1    | 1    | 1    | 1  | 1    | 1    | 1    | 0    | 0    | 1   | 1  | 1  | 0   | 1  | 1    |
| PO  | 0  | 1  | 1  | 0   | 0   | 0  |    | 0  | 1   | 1  | 0   | 0   | 1   | 0   | 0   | 0    | 1    | 0    | 0    | 0  | 0    | 0    | 0    | 0    | 0    | 0   | 0  | 0  | 1   | 0  | 0    |
| MT  | 0  | 1  | 1  | 0   | 1   | 1  | 0  |    | 1   | 1  | 0   | 1   | 1   | 0   | 0   | 1    | 1    | 0    | 0    | 0  | 0    | 0    | 0    | 0    | 0    | 1   | 0  | 1  | 1   | 0  | 0    |
| V4t | 0  | 1  | 1  | 0   | 1   | 1  | 1  | 1  |     | 1  | 0   | 1   | 1   | 0   | 0   | 1    | 1    | 0    | 1    | 1  | 1    | 0    | 1    | 0    | 0    | 1   | 1  | 1  | 1   | 0  | 0    |
| DP  | 0  | 1  | 1  | 0   | 1   | 1  | 1  | 0  | 1   |    | 0   | 1   | 1   | 0   | 0   | 1    | 1    | 0    | 1    | 1  | 1    | 0    | 1    | 0    | 0    | 1   | 1  | 1  | 1   | 0  | 0    |
| VOT | 0  | 0  | 0  | 0   | 0   | 0  | 0  | 0  | 0   | 0  |     | 0   | 0   | 0   | 0   | 0    | 0    | 0    | 0    | 0  | 0    | 0    | 0    | 0    | 0    | 0   | 0  | 0  | 0   | 0  | 0    |
| LIP | 0  | 1  | 1  | 0   | 1   | 1  | 0  | 1  | 1   | 1  | 0   |     | 1   | 0   | 0   | 1    | 1    | 0    | 0    | 0  | 0    | 0    | 0    | 0    | 0    | 1   | 0  | 1  | 1   | 0  | 0    |
| VIP | 0  | 1  | 1  | 0   | 1   | 1  | 1  | 1  | 1   | 1  | 0   | 1   |     | 0   | 0   | 1    | 1    | 0    | 1    | 1  | 1    | 0    | 1    | 0    | 0    | 1   | 1  | 1  | 1   | 0  | 0    |
| MIP | 0  | 0  | 0  | 0   | 0   | 0  | 0  | 0  | 0   | 0  | 0   | 0   | 0   |     | 1   | 0    | 0    | 0    | 0    | 0  | 0    | 0    | 0    | 0    | 0    | 0   | 0  | 0  | 0   | 0  | 0    |
| MDP | 0  | 0  | 0  | 0   | 0   | 0  | 0  | 0  | 0   | 0  | 0   | 0   | 0   | 1   |     | 0    | 0    | 0    | 0    | 0  | 0    | 0    | 0    | 0    | 0    | 0   | 0  | 0  | 0   | 0  | 0    |
| MSTd| 0  | 1  | 1  | 0   | 1   | 1  | 0  | 1  | 1   | 1  | 0   | 1   | 1   | 0   | 0   |      | 1    | 0    | 1    | 1  | 0    | 1    | 0    | 0    | 0    | 1   | 1  | 1  | 1   | 0  | 0    |
| MSTl| 0  | 1  | 1  | 0   | 1   | 1  | 1  | 1  | 1   | 1  | 0   | 1   | 1   | 0   | 0   | 1    |      | 0    | 1    | 1  | 1    | 0    | 1    | 0    | 0    | 1   | 1  | 1  | 1   | 0  | 1    |
| PITd| 1  | 0  | 0  | 0   | 0   | 1  | 0  | 0  | 0   | 0  | 0   | 0   | 0   | 0   | 0   | 0    | 0    |      | 1    | 1  | 1    | 1    | 1    | 1    | 1    | 0   | 1  | 1  | 0   | 1  | 1    |
| PITv| 1  | 1  | 1  | 0   | 1   | 1  | 0  | 1  | 1   | 0  | 0   | 1   | 0   | 0   | 0   | 1    | 1    | 1    |      | 1  | 1    | 1    | 1    | 0    | 0    | 0    | 1   | 1  | 0  | 1   | 1  | 1    |
| 7a  | 1  | 1  | 1  | 0   | 1   | 1  | 0  | 0  | 1   | 1  | 0   | 1   | 0   | 0   | 1   | 1    | 1    | 1    | 1    |    | 1    | 1    | 0    | 0    | 0    | 1   | 1  | 0  | 1   | 1  | 1    |
| STPp| 1  | 1  | 0  | 0   | 0   | 1  | 0  | 1  | 0   | 1  | 0   | 1   | 0   | 0   | 0   | 1    | 1    | 1    | 1    | 1  |      | 1    | 1    | 1    | 0    | 1   | 1  | 0  | 1   | 0  | 1    |
| CITd| 1  | 0  | 0  | 0   | 0   | 1  | 0  | 0  | 0   | 0  | 0   | 0   | 0   | 0   | 0   | 0    | 0    | 1    | 1    | 1  | 1    |      | 1    | 1    | 1    | 0   | 1  | 1  | 0   | 1  | 1    |
| CITv| 1  | 1  | 0  | 0   | 0   | 1  | 0  | 0  | 1   | 0  | 0   | 1   | 0   | 0   | 0   | 1    | 0    | 1    | 1    | 1  | 1    | 1    |      | 1    | 1    | 0   | 1  | 1  | 0   | 1  | 1    |
| STPa| 1  | 0  | 0  | 0   | 0   | 0  | 0  | 0  | 0   | 0  | 0   | 0   | 0   | 0   | 0   | 0    | 0    | 1    | 0    | 0  | 1    | 1    | 1    |      | 1    | 0    | 0  | 0  | 0   | 1  | 1    |
| AITv| 1  | 0  | 0  | 0   | 0   | 0  | 0  | 0  | 0   | 0  | 0   | 0   | 0   | 0   | 0   | 0    | 1    | 0    | 0    | 1  | 1    | 1    | 1    | 1    |      | 0   | 0  | 0  | 0   | 1  | 1    |
| FEF | 0  | 1  | 1  | 0   | 1   | 1  | 0  | 1  | 1   | 0  | 1   | 1   | 0   | 0   | 1   | 1    | 0    | 0    | 0    | 0  | 0    | 0    | 0    | 0    | 0    |     | 1  | 1  | 0   | 0  | 0    |
| TF  | 1  | 1  | 1  | 0   | 1   | 1  | 0  | 0  | 1   | 0  | 0   | 1   | 0   | 0   | 0   | 1    | 1    | 1    | 1    | 1  | 1    | 1    | 1    | 0    | 0    | 1   |    | 1  | 0   | 1  | 1    |
| 46  | 1  | 1  | 1  | 0   | 1   | 1  | 0  | 1  | 1   | 1  | 0   | 1   | 0   | 0   | 1   | 1    | 1    | 1    | 1    | 1  | 1    | 1    | 0    | 0    | 1    | 1   | 1  |    | 0   | 1  | 1    |
| FST | 0  | 1  | 1  | 0   | 1   | 0  | 1  | 1  | 1   | 1  | 0   | 1   | 0   | 0   | 1   | 1    | 0    | 0    | 0    | 0  | 0    | 0    | 0    | 0    | 0    | 0   | 0  | 0  |     | 0  | 0    |
| TH  | 1  | 0  | 0  | 0   | 0   | 1  | 0  | 0  | 0   | 0  | 0   | 0   | 0   | 0   | 0   | 0    | 0    | 1    | 1    | 1  | 1    | 1    | 1    | 1    | 1    | 0   | 1  | 1  | 0   |    | 1    |
| AITd| 1  | 0  | 0  | 0   | 0   | 1  | 0  | 0  | 0   | 0  | 0   | 0   | 0   | 0   | 0   | 0    | 1    | 1    | 1    | 1  | 1    | 1    | 1    | 1    | 1    | 0   | 1  | 1  | 0   | 1  |      |

#### Figure 8
Confirmation or mismatch of connections, and prediction of unknown connections in a reconstructed submatrix of the cortical network. Data for the reconstruction of this 31 × 31 graph was based on ref [9]. Green fields denote confirmed existing (1) and absent (0) connections, respectively, whereas red fields indicate a mismatch between the original and the shown reconstructed connectivity (either by inserting connections into the matrix or removing them from the original). Yellow fields highlight connections that were predicted to exist (1) or to be absent (0) by the reconstruction approach and whose status was previously not known.

$$k_i = \sum_{j=1}^{N} K_{non}(i,j) = \sum_{j=1}^{N} K_{non}(j,i) \qquad (2)$$

Note that the node degree provides a direct measurement of the degree in which the specific node is connected to the rest of the network.

*Feature 2 (Topological) – Clustering coefficient*
Given a subset $S$ of the network nodes, the clustering coefficient of this set [25] can be defined as the ratio between the number of edges between the elements of $S$ and the maximum possible number of such connections. Therefore, the *clustering coefficient* of a specific node $i$ can be more formally defined as

$$CC_i = 2 \frac{E(W_i)}{|W_i||W_i - 1|} \qquad (3)$$

where $W_i$ is the set containing the immediate neighbors of $i$, $E(W_i)$ is the number of edges between such neighbors, and $|W_i|$ is the number of elements in the set $S$. The clus-

**Table 8: Potential impact of unknown data.**

| Version of matrix | Hamming error | Geometrical error |
|---|---|---|
| Original | 307 | 0.655 |
| Relative | 356 | 0.612 |
| Full | 278 | 0.731 |

Assessment of reconstructions for a 31 × 31 subgraph, based on data from ref. [9], which either was considered in its 'original' state, with all currently absent connections assumed to exist ('full'), or with about 1/3 of the absent connections assumed to exist ('relative').

tering coefficient of node $i$ therefore expresses how intensely interconnected the neighbors of the reference nodes are concerning direct connections between the nodes. Note that $0 \leq C C_i \leq 1$.

*Feature 3 (Topological) – Average shortest path distance*
Given any two nodes $i$ and $j$ of a network, they are said to be connected in case there is a sequence of edges extending from one of those nodes to the other, possibly passing through several relay nodes. The *shortest path* $s_{i,j}$ between the two nodes $i$ and $j$ corresponds to the path involving the smallest sum of involved edge segments. The *shortest path distance* is defined as being equal to the respective sum of edge segments. Note that the shortest path may not be unique, but all such paths will have the same shortest path distance. Because the shortest path is defined with respect to a pair of nodes, and we want to assign a related measurement to each node $i$, we henceforth consider the average of the shortest paths between $i$ and all the other network nodes as a feature of node $i$, represented as $s_i$. Note that all nodes in the cortical network are connected.

*Feature 4 (Topological) – Matching index*
This measurement, introduced in [26,27] applies to any pair of nodes $i$ and $j$ (connected or not) and can be conceptually defined as the amount of connectivity overlap between each of those nodes and the remainder of the network. More specifically, in case of non-directed networks, $|a_i \wedge a_j|$ is the number of common projections that occur in nodes $i$ as well as $j$ denoting the number of common target or source nodes for projection fibers. The total number of connections that occur in node $i$, in node $j$, or in both nodes is denoted as $|a_i \vee a_j|$. The matching index is then calculated as:

$$m_{i,j} = \frac{|a_i \wedge a_j|}{|a_i \vee a_j|} \quad (4)$$

A low matching index value indicates that the nodes have diverging input and output and are linked to substantially different parts of the network. As with the shortest path, the matching indices are averaged for all nodes.

*Feature 5 (Spatial) – Local density*
It is often the case with point distributions (as the centers of mass of the cortical areas) that the number of points per unit area varies along the space. In such cases, it is interesting to consider the *local density* around each point. This value can be estimated by dividing the number $P_i$ of neighboring points contained in a sphere of small radius $R$ centered at the reference point $i$ by the volume of that sphere, i.e.

$$L_i(R) = \frac{3}{4} \frac{P_i(R)}{\pi R^3} \quad (5)$$

The quantity $P_i(R)$ has been calculated with respect to each node $i$ by counting how many nodes are at distance smaller or equal to $R = 15$, which corresponds to about 1/4 of the maximum internode distances in the cortical networks. Note that this measurement is influenced by the volume of each cortical region. The larger the volume, the smaller the local density.

*Feature 6 (Spatial) – Coefficient of variation of the nearest distances*
Given a reference point $i$ and a maximum radius $R$, the nearest neighbors $Q$ of that point can be defined as those points which are contained in the sphere of radius $R$ center at point $i$. The measurements in this work assumed $R = 15$. The *nearest distances* of point $i$ are therefore defined as the set of the Euclidean distances between it and each of the nearest neighboring nodes in $Q$. The coefficient of variation (i.e. the standard deviation divided by the average) of the nearest distances provides an interesting indication about the local distance regularity around each reference point. For instance, a low value of this coefficient indicates that the nearest neighbors of a point are almost equidistant. As with the previous measurement, the coefficient of variation of the nearest distances can also be affected by the volume of the cortical region. More specifically, the larger the volume, the larger this measurement tends to be.

*Feature 7 (Spatial) – Area size of each cortical region*
This measurement corresponds simply to the area size of the two-dimensional surface of each cortical region. The surface area was measured directly within three-dimensional space; that means, we did not use a flattened two-dimensional map to estimate the surface extent of a cortical region.

*Feature 8 (Spatial) – Cartesian coordinates of the cortical areas center of mass*
These features, considered together for simplicity's sake, correspond to the *x*, *y* and *z* coordinates of the center of mass of each cortical area. By application of the wiring rule described below to this feature, network nodes were linked that are spatially close to each other.

Table 9 summarizes the eight measurements considered in this work and their respective identifications.

### Network reconstruction from node features
Hypothetical cortical networks were created by assessing the pairwise similarity of nodes with respect to each of the eight features. Undirected links were created between nodes, if their similarity exceeded a threshold. In order to avoid the need to specify this threshold, we considered a sequence of equally spaced thresholds during the reconstruction and took as result the threshold leading to the best results (i.e., best recovery of the original connectivity).

The topological and spatial context around each node $i$ in the two communities can be characterized in terms of respective feature vector $\vec{v}_i$ containing a subset of selected measurements at that node. In this work, we consider 1-by-1, 2-by-2 and 3-by-3 combinations of the six measurements described in the section 'Network characterization indices' above. In order to avoid bias implied by the different ranges of each measurement, these values have been standardized [28]. More specifically, for each type of measurement, each value was subtracted from the respective average and divided by the respective standard deviation. Note that these new measurements have zero average and unit standard deviation.

**Table 9: Node-based characterization measures used for network reconstruction.**

| MEASUREMENTS | IDENTIFICATION NUMBER |
|---|---|
| **Node degree** | 1 |
| **Clustering coefficient** | 2 |
| **Avg. shortest path distance** | 3 |
| **Matching index** | 4 |
| *Local density* | 5 |
| *Standard deviation of nearest distances* | 6 |
| *Area size* | 7 |
| *Cartesian coordinates (x, y, z)* | 8 |

The table lists node-based measurements for network reconstruction by considering topological (in bold) and spatial (in italics) properties of the cortical data.

Now it is possible to use the methodology suggested in [29,30] in order to obtain networks from the feature vectors $\vec{v}_i$. Note that each element $v_i(p)$ in such a vector represents a possible measurements. In order to do so, we start with $N = 95$ isolated nodes and, for each possible pair of nodes $(i, j)$, we establish a connection between them, by making and $K(i,j) = 1$ and $K(j,i) = 1$, whenever the following condition is met

$$d(i,j) = \sqrt{\sum_{p=1}^{n} \left(v_i(p) - v_j(p)\right)^2} \leq T \qquad (6)$$

where $n$ is the number of chosen measurements and $T$ is a pre-fixed threshold. Note that the smaller the value of this threshold, the less intensely connected the respective network will result.

### Network comparisons
Because the reconstructed networks are fully congruent with the original data, in the sense that they have the same number of nodes and each node refer to the same cortical region, it is possible to obtain a simple and effective measurement of the difference between the original network $G$ and each of the networks $F$ obtained from the topological and spatial features in terms of the distance defined as being equal to the number of different entries in the respective adjacency matrices. More formally, we have that

$$H(G,F) = \frac{1}{2}\left\{ N^2 - \sum_{i=1}^{N}\sum_{j=1}^{N} \delta\left(K_G(i,j), K_F(i,j)\right) \right\} \qquad (7)$$

where $K_G$ and $K_F$ are the adjacency matrices of the original and reconstructed matrices, and $\delta(a, b)$ is the Kronecker delta function, which results 1 whenever $a$ and $b$ are equal and 0 otherwise. The 1/2 factor is necessary in order to account for the fact that in a non-directed graph each edge appears twice in the respective adjacency matrix.

However, this measure, the Hamming distance, provides a biased quantification of the similarity between any two matrices in case the number of zeros and ones is significantly different. For instance, in case a matrix contains few ones and many zeros, its Hamming distance to a null matrix (all entries equal to zero) will be very small. In order to provide a more balanced overall measurement of the similarity between two adjacency matrices $A$ and $B$, both with the same dimension $N \times N$, we consider the geometrical average $g(A, B)$ between the ratios of correct ones ($R_1$) and correct zeros ($R_0$). More specifically, in case matrix $A$ contains $A_0$ zeroes and $A_1$ ones, and matrix $B$ contains $b_0$ zeroes coinciding with the zeroes of $A$ and $b_1$ coinciding ones, we define $R_1 = b_1/A_1$ and $R_0 = b_0/A_0$. The two matrices will be maximally similar in case $g(A, B) = \sqrt{R_1 R_0} = 1$, which is verified if and only $R_1 = R_0 = 1$.

It is possible to obtain a random reference for the comparison between any two adjacency matrices $A$ and $B$ as follows. Let the ratio of ones in $A$ be $r_1 = A_1/N^2$ and the ratio of zeroes be $r_0 = A_0/N^2$. It can be shown that the average expected ratios of correct ones and zeros while comparing matrix $A$ with matrices $B$ generated randomly (uniform probability) with the same ratio $r_1$ of ones are given as $R_1 = r_1$ and $R_0 = r_0$.

These comparisons with the original connectivity and random benchmarks were applied to all adjacency matrices reconstructed from individual and combined node features.

## Authors' contributions

The initial proposal of checking the relationship between spatial position and connectivity was suggested by CCH and MK, while LDFC proposed the methodology of network reconstructions. LDFC performed all experimental simulations and analyses, except the determination of the correlation statistical tests, performed by CCH. The discussion and interpretation of the results, as well as the paper writing, was performed jointly by the three authors.

## Additional material

### Additional file 1
*Analysis of* C. elegans *data. The file provides the results of a supplementary data analysis of the neuronal network of* C. elegans, *using the same network reconstruction approach as for primate cortical connectivity.*
Click here for file
[http://www.biomedcentral.com/content/supplementary/1752-0509-1-16-S1.doc]

## Acknowledgements
Luciano da F. Costa thanks FAPESP (05/00587-5) and CNPq (308231/03-1) for sponsorship. Marcus Kaiser acknowledges support from EPSRC (EP/E002331/1).